\documentclass{PoS}

\title{
\vspace{-1cm}
\begin{minipage}{\textwidth}
\begin{flushright}
\normalsize ITEP-LAT/2007-18 \\ HU-EP-07/39\\
\end{flushright}
\end{minipage}\\[15pt]
Study of the topological vacuum structure of $SU(2)$ gluodynamics
at $T > 0$ with overlap fermions and improved action
\thanks{This work is supported by joint grants RFBR-DFG 06-02-04010, DFG-RFBR 436 RUS 113/739/0-2. V.G.B.,
S.M.M. and M.I.P. are supported by grants RFBR 05-02-16306, 07-02-00237-a and by the EU Integrated
Infrastructure Initiative Hadron Physics (I3HP) under contract RII3-CT-2004-506078.The work of E.-M. I.
is supported by DFG under contract FOR 465 / Mu932/2-4. S.M.M. is also supported by an INTAS~YS fellowship 05-109-4821.}}

\ShortTitle{Study of the topological
structure of $SU(2)$ gluodynamics
at $T > 0$}

\author{V.G.~Bornyakov~,\\
Institute of High Energy Physics, Protvino 142284, Moscow Region, Russia\,\,\, and \\
Institute of Theoretical and Experimental Physics, B.Cheremushkinskaya 25, Moscow 117259, Russia\\
E-mail: \email{Vitaly.Bornyakov@ihep.ru}}

\author{\speaker{E.V.~Luschevskaya}, S.M.~Morozov and M.I.~Polikarpov\\
Institute of Theoretical and Experimental Physics, B.Cheremushkinskaya 25, Moscow 117259, Russia\\
E-mail: \email{luschevskaya@itep.ru}\quad
        \email{smoroz@itep.ru}\quad
        \email{polykarp@itep.ru}}

\author{E.-M.~Ilgenfritz and M.~M\"uller-Preussker\\
Humboldt-Universit\"at zu Berlin, Institut f\"ur Physik, Newtonstr. 15, D-12489 Berlin, Germany\\
E-mail: \email{ilgenfri@physik.hu-berlin.de}\quad
        \email{mmp@physik.hu-berlin.de}}

\abstract{We study $SU(2)$ gluodynamics at finite temperature
near the deconfining phase transition.
We create the lattice ensembles using the tadpole improved L\"uscher-Weisz action.
The overlap Dirac operator is used to determine
the following three aspects of vacuum structure:
({\it i}) The topological susceptibility
is evaluated at various temperatures across the phase transition, ({\it ii})
the overlap fermion spectral density is determined and found to depend on the
Polyakov loop above the phase transition and ({\it iii}) the corresponding
localization properties of low lying eigenmodes are investigated.}

\FullConference{The XXV International Symposium on Lattice Field Theory\\
         July 30 - August 4, 2007\\
         Regensburg, Germany}

\begin{document}

\section{Introduction}

More than ten years ago, using a model generalizing random matrix theory,
M.A. Stephanov~\cite{Stephanov} predicted, that in $SU(3)$ gluodynamics
above $T_c$ the different Polyakov loop sectors behave differently.
In the complex-valued Polyakov loop sectors the chiral condensate should
turn to zero at $T$ substantially above $T_c$.
For $SU(2)$ lattice gluodynamics, where the Polyakov loop is real, it was
predicted
that the chiral condensate stays non-zero, $<\bar{\psi}\psi> \neq 0$, for all
temperatures $T > T_c$ in the sector with a negative averaged Polyakov loop $L<0$.

As for $SU(3)$, Gattringer {\it et al.}~\cite{Gattringer} came to a different conclusion. They defined a new observable,
the gap in the Dirac spectrum, and used it as an order parameter for the
restoration of chiral symmetry. It was found that the spectral gap opens
up at one single temperature $T=T_c$ in all three $Z_{3}$ sectors.
Here we examine whether Stephanov's prediction for the Dirac spectrum remains
valid in the case of $SU(2)$ gluodynamics in the deconfined phase.

\section{Improved action}

Ensembles of $O(100)$ statistically independent quenched $SU(2)$ configurations
are generated with the tadpole improved L\"uscher-Weisz action on $20^3\times6$
lattices.
This action is known to suppress dislocations. The form of the action is :
\begin{equation}
S=\beta_{imp}\sum_{pl}S_{pl}-\frac{\beta_{imp}}{20u_{0}^2}\sum_{rt}S_{rt} \; ,
\label{eq:action}
\end{equation}
where $S_{pl}$ and $S_{rt}$ denote the plaquette and $1x2$
rectangular loop terms in the action,
$S_{pl,rt}=\frac{1}{2}Tr(1-U_{pl,rt})$. The factor
$u_{0}=(W_{1x1})^{1/4}$ is the {\it input} tadpole factor. It is
determined from $W_{1x1}=\langle (1/2)TrU_{pl} \rangle$
computed at zero temperature \cite{universality}.
The deconfining phase transition occurs at
$\beta_{imp}=\beta_c=3.248(2)$ for $N_{\tau}=6$,
which corresponds to $T_{c}/\sqrt{\sigma}=0.71(2)$~\cite{calorons}.

\section{Massless overlap Dirac operator}

The massless overlap Dirac operator has the form~\cite{Neuberger}
\begin{equation}
D_{ov}=\frac{\rho}{a} \left( 1 + D_W \Big/ \sqrt{D^{\dagger}_W D_W} \right) \; ,
\label{eq:overlap}
\end{equation}
where $D_W = M - \rho/a$ is the Wilson Dirac operator with a negative mass
term, $M$ is the Wilson hopping term, $a$ is the lattice spacing. The optimal
value of the $\rho$ parameter is found to be $1.4$ also for the lattice
ensembles under investigation. Anti-periodic (periodic) boundary conditions in
time (space) directions are imposed to the fermionic field.

In order to compute the sign function
\begin{equation}
D_W \Big/ \sqrt{D^{\dagger}_W D_W} = \gamma_5~{\rm sgn}(H_W) \; ,
\label{eq:signfunction}
\end{equation}
where $H_W = \gamma_5~D_W$ is the {\it hermitian} Wilson Dirac operator,
we use the minmax polynomial approximation.
The overlap Dirac operator constructed this way preserves the chiral symmetry
even on the lattice and allows to study the properties of the Dirac modes from
first principles.
It will be called $D$ in the following and replaces the continuum Dirac operator
$D=D_{\mu}~\gamma_{\mu}$ where $D_\mu=\partial_\mu-igA_\mu$ is the covariant
partial derivative with the gauge field background $A_{\mu}$.

\section{Topological susceptibility $\chi_{top}(T)$}

\begin{figure}
\begin{tabular}[h]{cc}
\hspace{0.5cm}
\includegraphics[scale=0.335,angle=270]{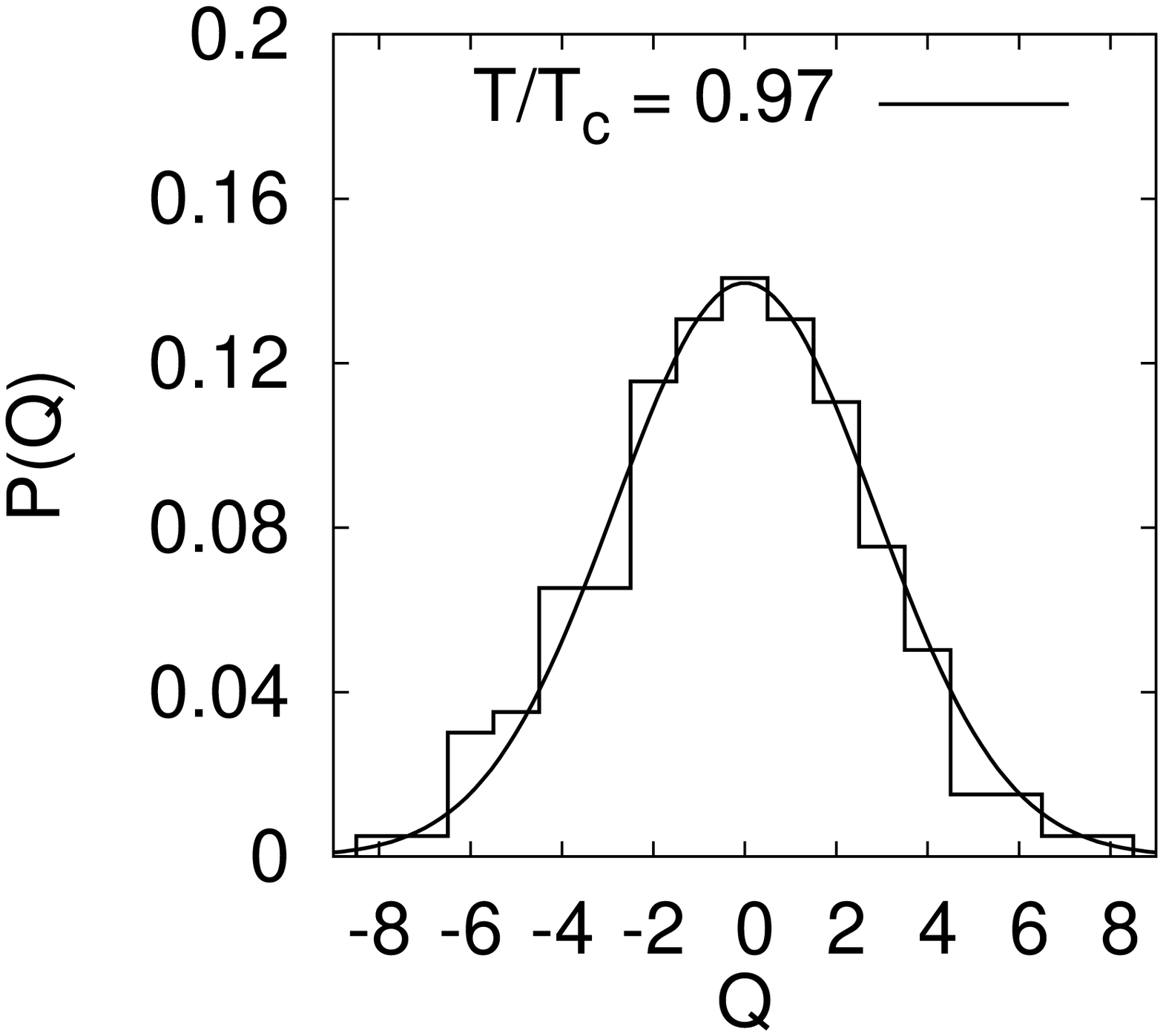} &
\includegraphics[scale=0.3,angle=270]{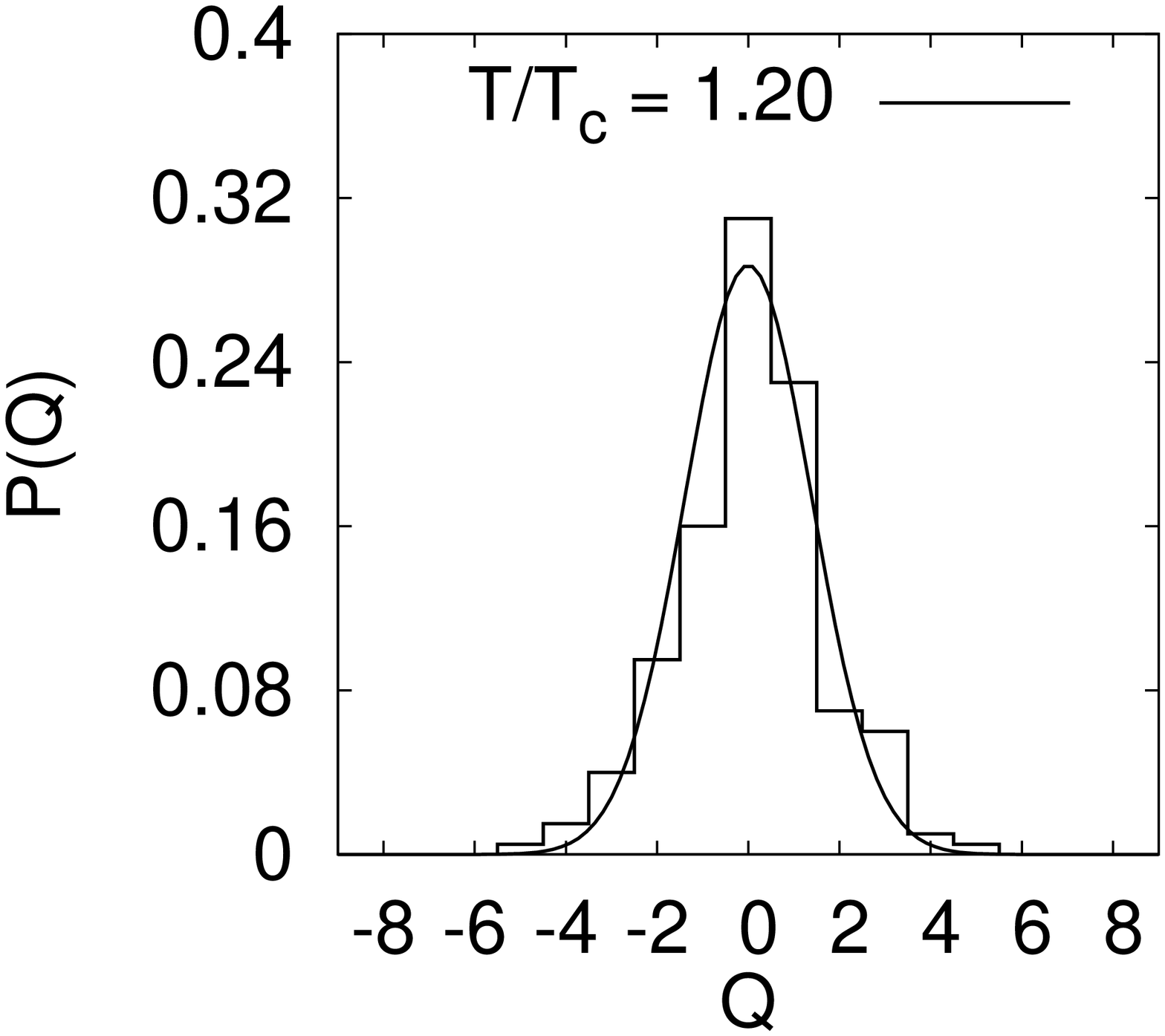}\\

\end{tabular}
\caption{Probability distributions of the topological charge $Q$ for
two temperatures below and above $T_c$.} \label{fig:fig1}
\end{figure}

We solved the Dirac equation numerically for its eigensystem
\begin{equation}
D~\psi_n = \lambda_n~\psi_n
\label{eq:eigensystem}
\end{equation}
and considered 50 lowest eigenvectors. As a first application we
search for the exact zero modes. Their number is related to the
total topological charge $Q_{top}$ of the lattice configuration
through the Atiyah-Singer index theorem :
\begin{equation}
Q_{top} = Q_{index} = N_{-} - N_{+} \; ,
\label{eq:indextheorem}
\end{equation}
where $N_{-}$ and $N_{+}$ are the numbers of fermionic modes with negative
and positive chirality $\psi^{\dagger}~\gamma_5~\psi$, respectively.
For the lattice ensembles the expectation
value $\langle Q_{top} \rangle$ should vanish, but $\langle Q_{top}^2 \rangle$
measures the strength of global topological fluctuations.
The topological susceptibility is
\begin{equation}
\chi_{top} \equiv \frac{\langle Q_{top}^2 \rangle}{V} \; ,
\label{eq:susceptibility}
\end{equation}
where $V$ is the four-dimensional lattice volume in physical units.
In Fig.~\ref{fig:fig1} (left) we see a histogram of the topological charge
in the confinement phase, close to the transition.
Fig.~\ref{fig:fig1} (right) shows the corresponding histogram for a temperature
higher up in the deconfinement phase.
Both histograms can be approximately fitted by Gaussian distributions.

\begin{figure}
\begin{tabular}[h]{cc}
\includegraphics[scale=0.29,angle=270]{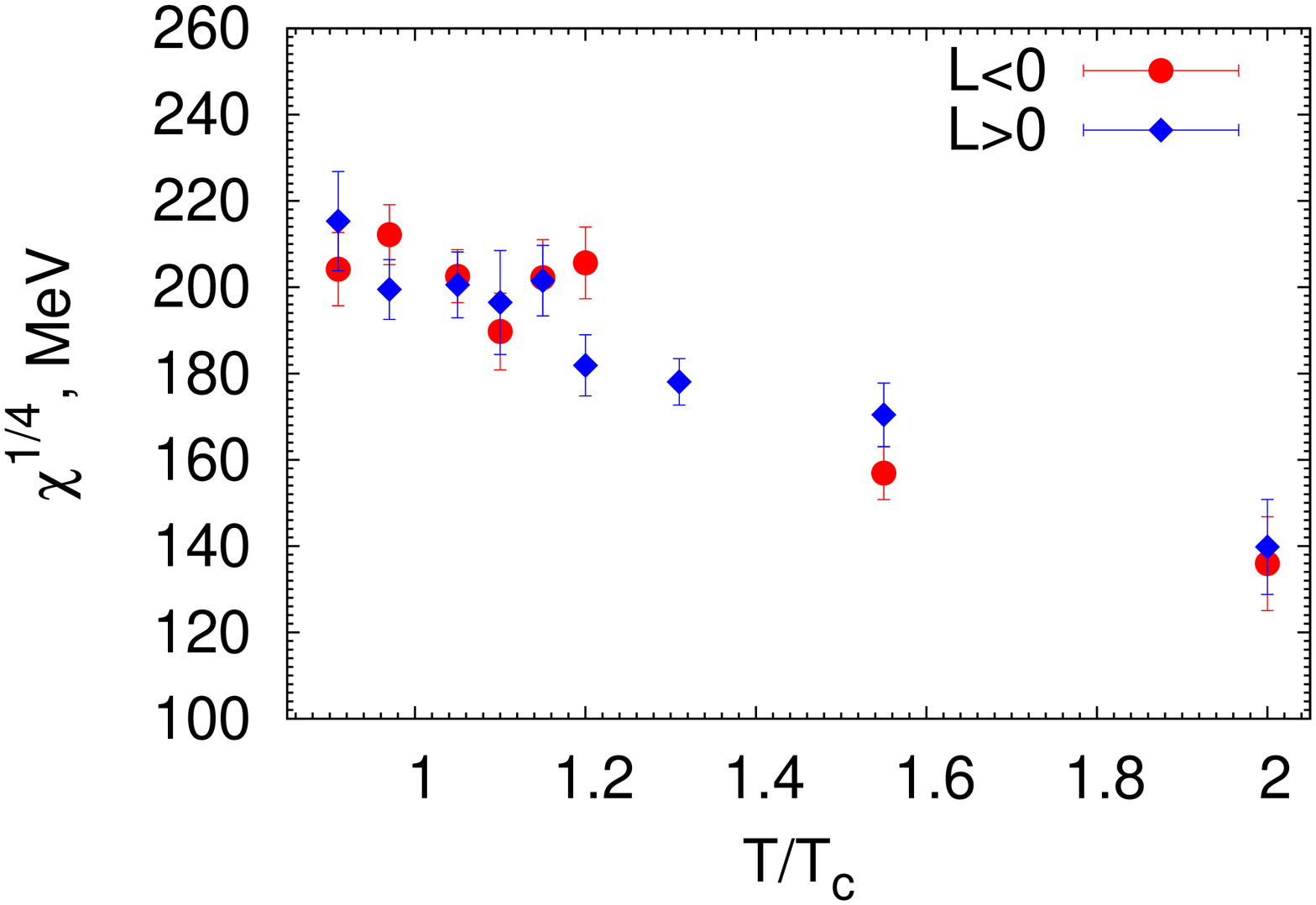}
&
\includegraphics[scale=0.29,angle=270]{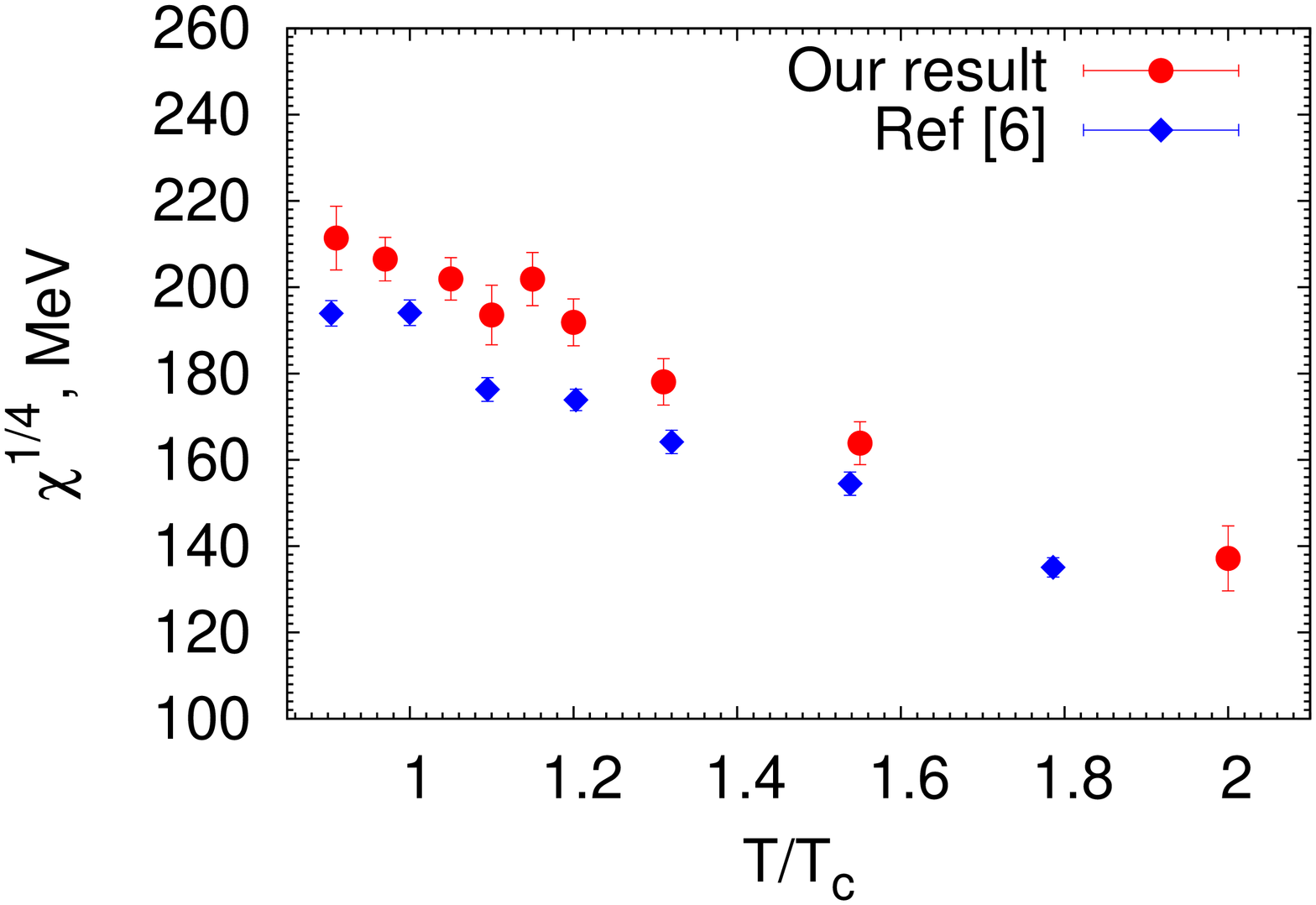}\\
\end{tabular}
\caption{The topological susceptibility $\chi_{top}$ as function of $T$
separately for $L > 0$ and $L < 0$ (left), and comparison of our
final result with that of Ref.~\cite{DiGiacomo} (right).}
\label{fig:fig2}
\end{figure}

Let us now discuss the topological susceptibility as function of temperature.
In Fig.~\ref{fig:fig2} (left)
we show, that the topological susceptibility $\chi_{top}$ in the negative
Polyakov loop sector ($L<0$) agrees at all $T$ within two standard deviations
with $\chi_{top}$ in the positive Polyakov loop sector ($L>0$).
In Fig.~\ref{fig:fig2} (right) we compare our final data for $\chi_{top}(T)$,
which are for $T<T_c$ averaged over all configurations and for $T>T_c$
only over the subsample with $L>0$,
with the results of Alles {\it et al.}~\cite{DiGiacomo}.
These authors presented the values of $10^{-4} \times \chi_{top}/\Lambda^{4}_L$
at various values of $\beta$ for Wilson's action representing different temperatures.
We took $\Lambda_{L}=14.15(42){\rm~MeV}$~\cite{DiGiacomo}
and extracted their susceptibility $\chi_{top}(T)$ from these data. The
topological susceptibility is slowly decreasing with
increasing temperature for both sets of data. Notice that the overlap definition
of $Q$ results in a
systematically higher susceptibility than the improved field theoretic
definition employed by the Pisa group.

\section{Spectral density, chiral symmetry restoration and different $Z_N$ sectors}

\begin{figure}
\centerline{\includegraphics[scale=0.29,angle=270]{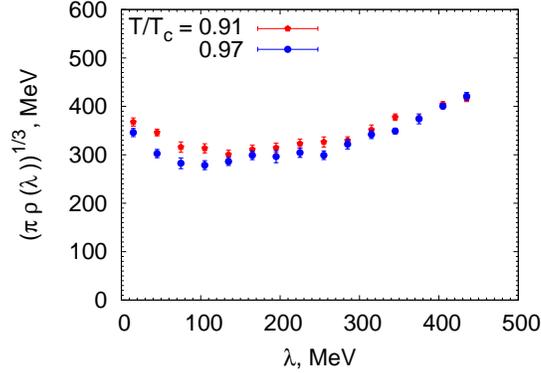}}
\caption{The spectral density of eigenmodes of the overlap Dirac
operator for two temperatures $T<T_c$ on the $20^3\times6$ lattice.}
\label{fig:fig3}
\end{figure}

The chiral condensate $\langle \bar{\psi}\psi \rangle$ is related to the density
$\rho(\lambda)$ of the non-zero eigenvalues $\lambda$ at $\lambda \rightarrow 0$
via the  Banks-Casher~\cite{BanksCasher} relation:
\begin{equation}
\langle \bar{\psi}\psi \rangle =- \lim_{\lambda \rightarrow 0}~\lim_{V \rightarrow \infty}~{\frac{\pi\rho(\lambda)}{V}} \; .
\label{eq:BanksCasher}
\end{equation}
The non-zero modes are globally non-chiral, but the near-zero ones are still locally chiral and correlated with lumps of the topological charge density. The number of modes belonging to this near-zero band is proportional to the total volume $V$.
In the chirally broken phase the required limit (\ref{eq:BanksCasher})
of $\rho(\lambda)$ is non-vanishing at $\lambda=0$~\cite{BanksCasher}.
In the chirally symmetric phase one expects $\rho(\lambda)=0$
in a finite region around the origin, {\it i.e.} that the spectrum develops a
gap. For the confinement (chirally broken) phase we find indeed that
the spectral density in physical units is practically constant
(almost $T$ independent), as can be seen in Fig.~\ref{fig:fig3}.
Comparing results for configurations with $L>0$ and $L<0$ we found that
at low $\lambda$ the density
$\rho(\lambda)$ is
$50 \sim 70$ MeV
higher for negative Polyakov loop sector.
We believe that this difference disappears in the thermodynamic limit.

\begin{figure}
\begin{tabular}[h]{cc}
  $~~~~~~~~~~L>0$ &  $~~~~~~~~~~L<0$ \vspace{-0.5cm}\\
\includegraphics[scale=0.29,angle=270]{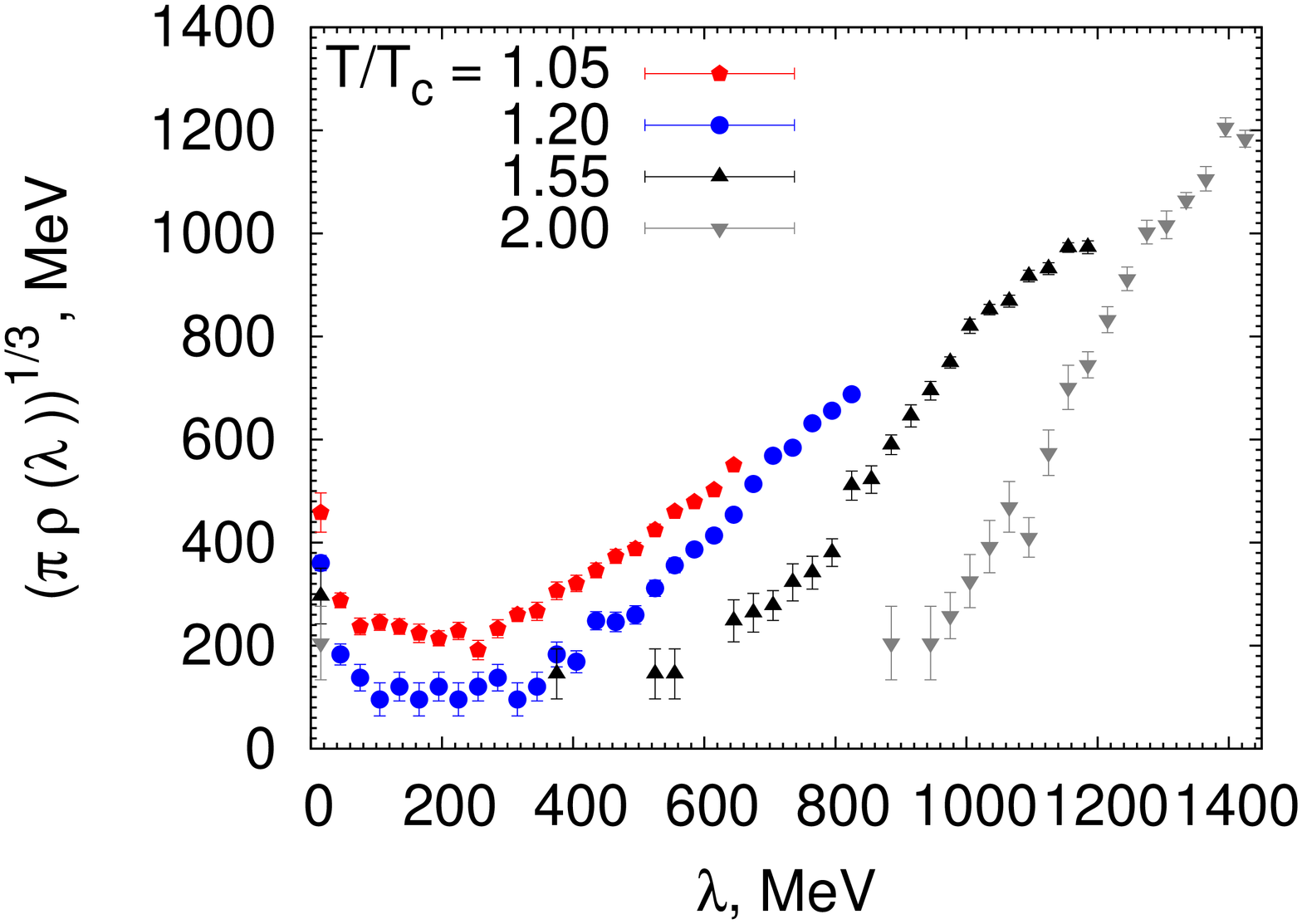}
&
\includegraphics[scale=0.29,angle=270]{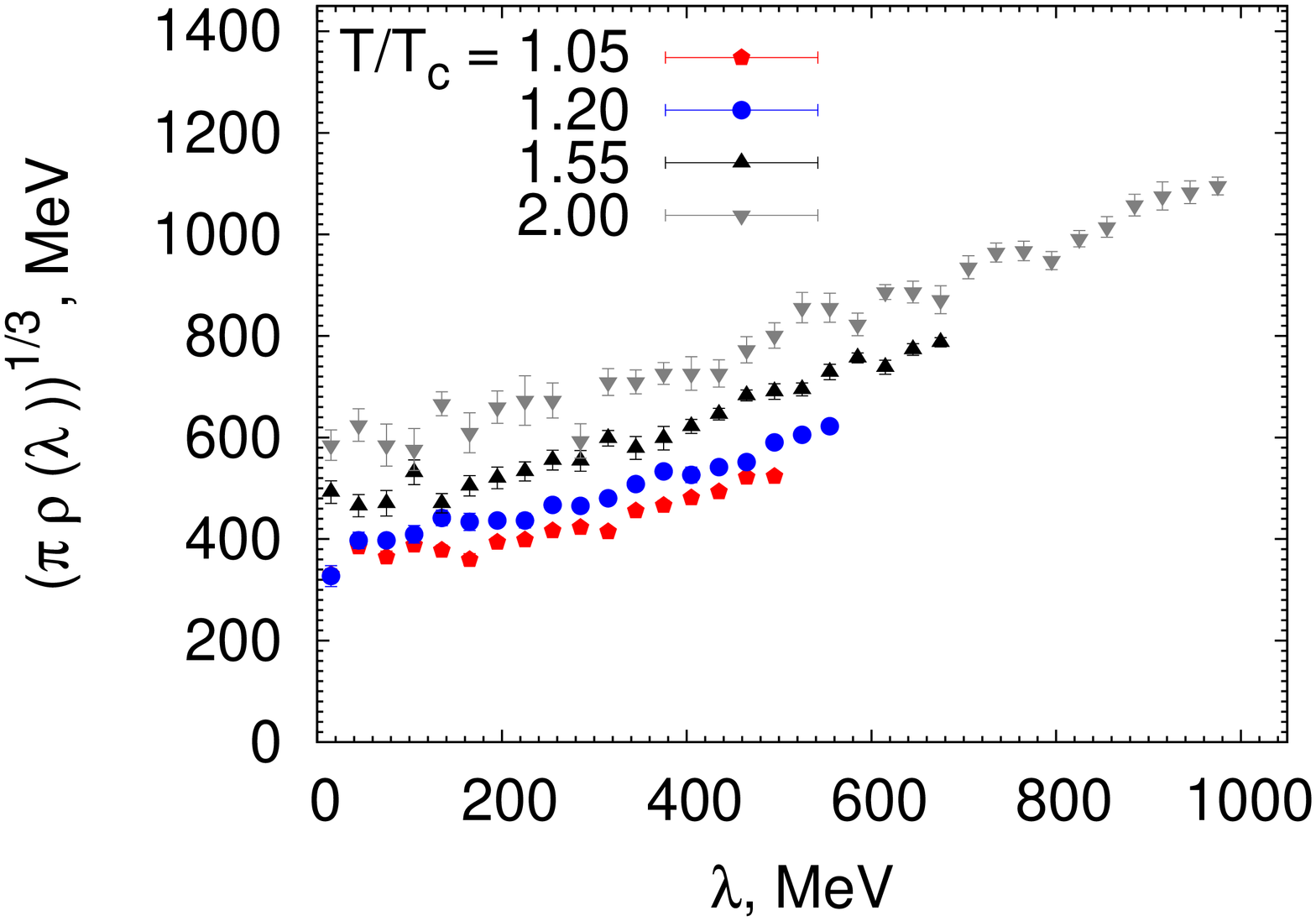}
\end{tabular}
\caption{The spectral density of eigenmodes of the overlap Dirac
operator for four temperatures $T>T_{c}$ on a $20^3\times6$ lattice,
evaluated separately according to the sign of the averaged Polyakov loop.}
\label{fig:fig4}
\end{figure}

For the deconfinement phase, when we take only configurations
with an average Polyakov loop $L>0$, Fig.~\ref{fig:fig4} (left) shows that
$\rho(\lambda)$ gradually decreases with increasing temperature, indicating
the decrease of the chiral condensate until a gap opens and gets wider.
For configurations with $L<0$  Fig.~\ref{fig:fig4} (right) shows
that $\rho(\lambda)$
at low lambda is nonzero and even grows with increasing temperature.

\vspace{-0.1cm}
\section{Spectral gap}
\vspace{-0.1cm}

\begin{figure}
\center{
\includegraphics[scale=0.35,angle=270]{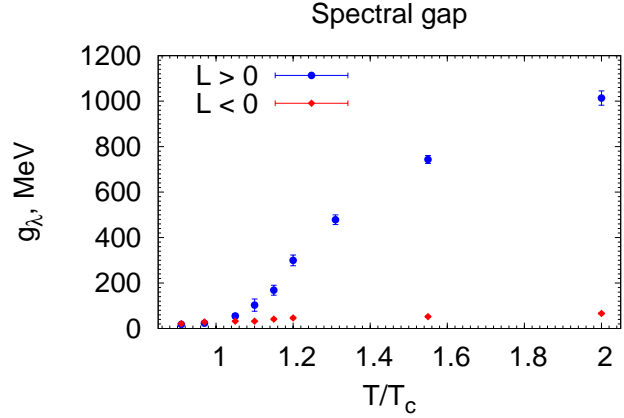}
\caption{The spectral gap for $SU(2)$ lattice gauge theory as
function of temperature, evaluated separately according to the sign
of the averaged Polyakov loop.} \label{fig:fig5}}
\end{figure}

The spectral gap $g_{\lambda}$
was defined by the smallest eigenvalue, which does not belong to a zero-mode.
In Ref.~\cite{Gattringer} Gattringer {\it et al.}
have shown for $SU(3)$ gluodynamics
that the gap, as function of temperature, has a similar behavior for the real
and both complex sectors corresponding to the phase of the averaged Polyakov loop.
The phase transition occurs at the same $T_c$, and with increasing lattice
volume the gap is decreasing.
Analogously, $SU(2)$ gluodynamics has only two sectors in the deconfinement phase,
distinguished by the sign of the (real-valued) averaged Polyakov loop.
We show in Fig.~\ref{fig:fig5}
a clearly defined and rapidly growing gap for configurations
with $L>0$, whereas for configurations with $L<0$ the gap remains very small up
to temperatures several times higher than $T_c$. The small gap
is a finite-volume effect and can be made to vanish in the limit
of spatial $V_3 \rightarrow \infty$.

\vspace{-0.1cm}
\section{Localization in different parts of the spectrum}
\vspace{-0.1cm}

\begin{figure}
\centerline{\includegraphics[scale=0.29,angle=270]{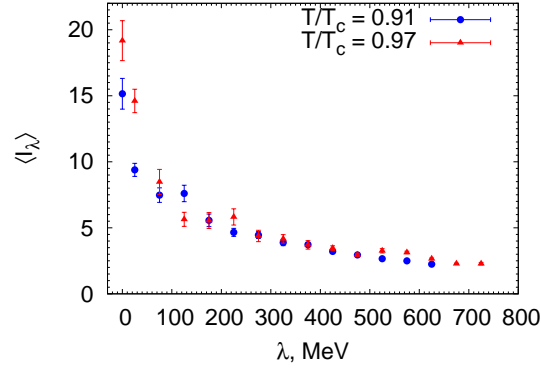}}
\caption{The average IPR within spectral bins for one temperature $T<T_{c}$.}
\label{fig:fig6}
\end{figure}

\begin{figure}
\begin{tabular}[h]{cc}
 $~~~~~~~~~~~~~~L>0$ &  $~~~~~~~~~~~~~~L<0$ \vspace{-0.5cm} \\
\includegraphics[scale=0.29,angle=270]{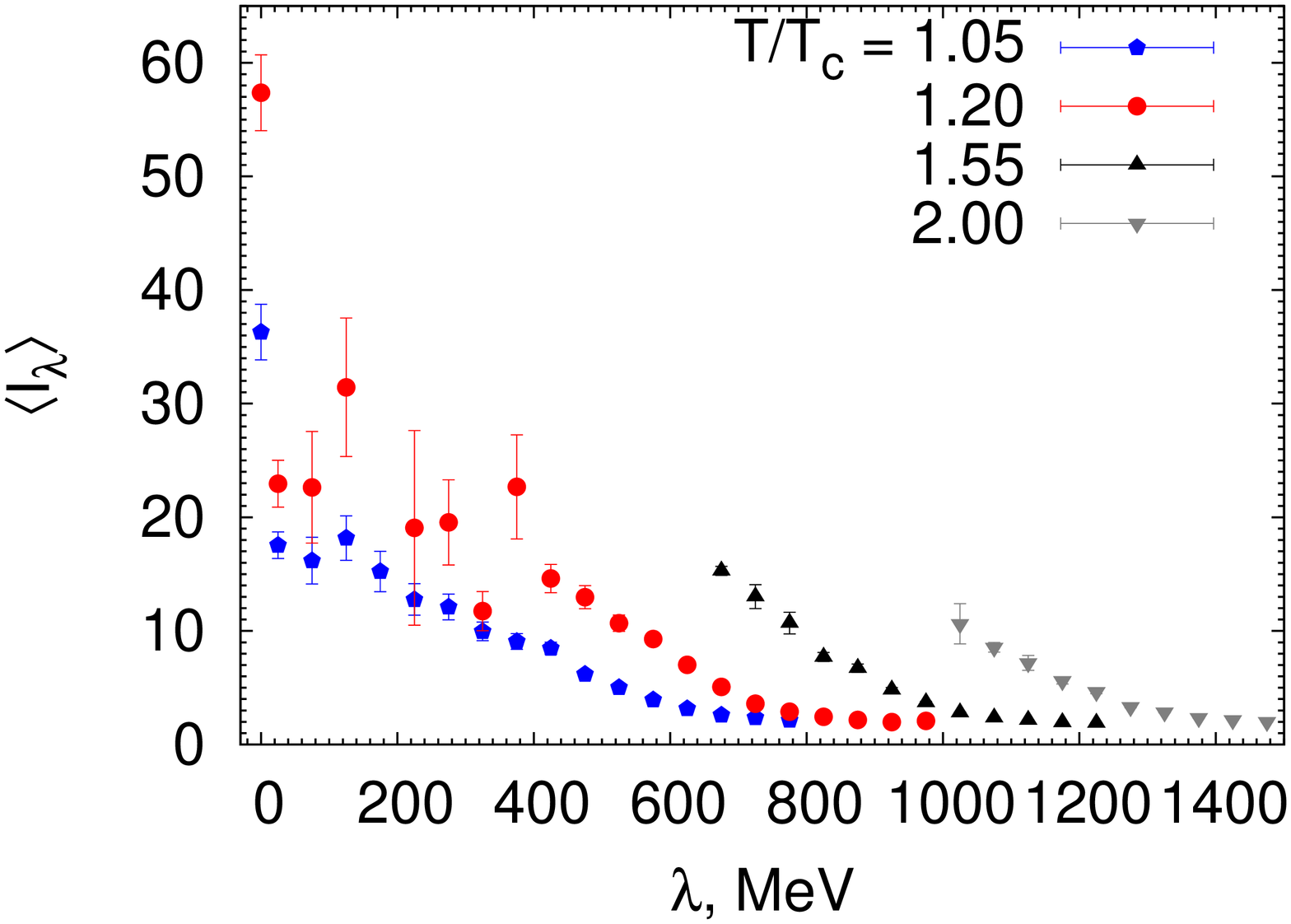}
&
\includegraphics[scale=0.29,angle=270]{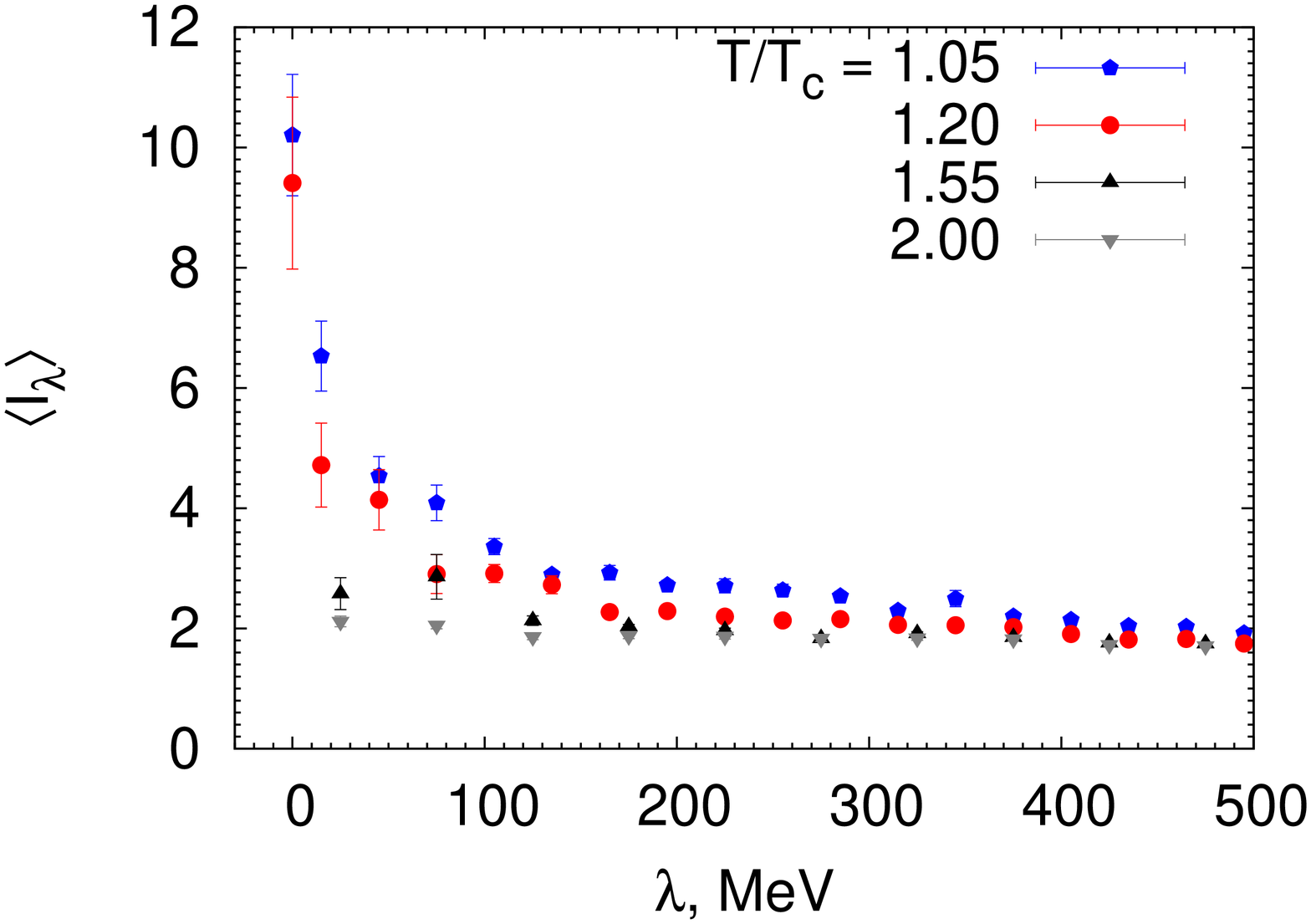}
\end{tabular}
\caption{The average IPR within spectral bins for four temperatures $T>T_{c}$,
separated according to the sign of the averaged Polyakov loop $L$.}
\label{fig:fig7}
\end{figure}

The scalar density of an eigenmode $\psi_{\lambda}(x)$ corresponding to an
eigenvalue $\lambda$ is denoted as
$\rho_{\lambda}(x)=\psi_{\lambda}^{\dagger}(x)\psi_{\lambda}(x)$, such that
$\sum_{x}\rho_{\lambda}(x)=1$
by virtue of normalization. The inverse participation ratio (IPR) $I_\lambda$ is the
natural measure of the localization. For any finite volume $V$ it is
defined by
\begin{equation}
I_{\lambda}=V\sum_{x}\rho_{\lambda}^2(x) \; .
\end{equation}
The IPR characterizes the inverse volume fraction of sites forming
the support of $\rho_{\lambda}(x)$. From Fig.~\ref{fig:fig6}
we conclude that for the temperature near but below $T_c$ the IPR
(localization) monotonously increases with decreasing eigenvalue.
There is no clear mobility edge.~\footnote{For $T=0$ the
localization of overlap eigenmodes has been investigated in
Refs.~\cite{Gubarev,Ilgenfritz}.} The monotony is not perfect among
the lowest one or two bins.
Thus, out of the low lying modes, the higher ones are continuously less localized.
We found that at these temperatures for configurations with negative
Polyakov loop, $L<0$, the modes are
by factor $2 \sim 3$ less localized. Again, as for $\rho(\lambda)$, this difference
should disappear in the thermodynamic limit.
In Fig.~\ref{fig:fig7} we show that with increasing temperature the average IPR
within the respective eigenvalue bins is increasing. The effect sets in for higher
and higher eigenvalues corresponding to the mobility edge moving outward in the deconfinement phase with the gap for $L>0$.
In the negative Polyakov loop sector the IPR is
constant at a low level, except for $\lambda < 500 {\rm~MeV}$, where the tendency
of the IPR to grow exists but is very weak.

\vspace{-0.1cm}
\section{Summary}
\vspace{-0.1cm}

We performed  first measurements of the topological
susceptibility with the help of the overlap Dirac operator
in finite temperature $SU(2)$ gluodynamics. We found that
the topological susceptibility
in the confinement phase is almost constant and is slowly decreasing
in the deconfinement phase, in agreement with previous results~\cite{DiGiacomo}.
We did not find systematic effects of the sign of the averaged Polyakov
loop on the topological susceptibility.
The chiral condensate, however, behaves completely different in the $L<0$ sector.
Chiral symmetry remains broken, the spectral gap stays close to zero
for all $T>T_{c}$ in agreement with Stephanov's model predictions.
A microscopic explanation in terms of the interplay of holonomy
and topology~\cite{calorons} needs to be worked ot.
This difference is accompanied by a different localization behavior
of the lowest fermionic eigenmodes in the two sectors.

\end{document}